\documentstyle[12pt]{article}

\newcommand{\infig}[2]{\begin{center}\mbox{\epsfxsize #2
\epsfbox{#1}}\end{center}}

\begin{document}
\input {epsf}

\newcommand{\beq}{\begin{equation}}
\newcommand{\eeq}{\end{equation}}
\newcommand{\beqa}{\begin{eqnarray}}
\newcommand{\eeqa}{\end{eqnarray}}

\def\ov{\overline}
\def\onlyif{\rightarrow}

\def\openone{\leavevmode\hbox{\small1\kern-3.8pt\normalsize1}}

\def\a{\alpha}
\def\b{\beta}
\def\g{\gamma}
\def\r{\rho}
\def\minus{\,-\,}
\def\eks{\bf x}
\def\kay{\bf k}

\def\ket#1{|\,#1\,\rangle}
\def\bra#1{\langle\, #1\,|}
\def\braket#1#2{\langle\, #1\,|\,#2\,\rangle}
\def\proj#1#2{\ket{#1}\bra{#2}}
\def\expect#1{\langle\, #1\, \rangle}
\def\trialexpect#1{\expect#1_{\rm trial}}
\def\ensemblexpect#1{\expect#1_{\rm ensemble}}
\def\kpsi{\ket{\psi}}
\def\kphi{\ket{\phi}}
\def\bpsi{\bra{\psi}}
\def\bphi{\bra{\phi}}

\def\ditto{\rule[0.5ex]{2cm}{.4pt}\enspace}
\def\th{\thinspace}
\def\ni{\noindent}
\def\thirty{\hbox to \hsize{\hfill\rule[5pt]{2.5cm}{0.5pt}\hfill}}

\def\set#1{\{ #1\}}
\def\setbuilder#1#2{\{ #1:\; #2\}}
\def\Prob#1{{\rm Prob}(#1)}
\def\pair#1#2{\langle #1,#2\rangle}
\def\Id{\bf 1}

\def\dee#1#2{\frac{\partial #1}{\partial #2}}
\def\deetwo#1#2{\frac{\partial\,^2 #1}{\partial #2^2}}
\def\deethree#1#2{\frac{\partial\,^3 #1}{\partial #2^3}}

\newcommand{\xx}{{\scriptstyle -}\hspace{-.5pt}x}
\newcommand{\yy}{{\scriptstyle -}\hspace{-.5pt}y}
\newcommand{\zz}{{\scriptstyle -}\hspace{-.5pt}z}
\newcommand{\kk}{{\scriptstyle -}\hspace{-.5pt}k}
\newcommand{\sx}{{\scriptscriptstyle -}\hspace{-.5pt}x}
\newcommand{\sy}{{\scriptscriptstyle -}\hspace{-.5pt}y}
\newcommand{\sz}{{\scriptscriptstyle -}\hspace{-.5pt}z}
\newcommand{\sk}{{\scriptscriptstyle -}\hspace{-.5pt}k}

\def\openone{\leavevmode\hbox{\small1\kern-3.8pt\normalsize1}}

\title{Quantum Cryptography using larger alphabets}
\author{
H. Bechmann-Pasquinucci and W. Tittel
\\
\small
{\it Group of Applied Physics, University of Geneva, CH-1211, Geneva 4,
Switzerland}}
\date{October 22, 1999}
\maketitle

\abstract{Like all of quantum information theory, quantum cryptography is   
traditionally based on two level quantum systems. In this letter, 
a new protocol for quantum key distribution 
based on higher dimensional systems is presented. 
An experimental 
realization using an interferometric setup is also proposed.  
Analyzing this protocol from the practical side, one
finds an increased key creation rate while keeping the initial laser pulse
rate constant.
Analyzing it for the case of intercept/resend eavesdropping strategy, 
an increased error rate is found compared to two dimensional systems, hence an
advantage 
for the legitimate users to detect an eavesdropper.} 
\vspace{1 cm}
\normalsize

\section {Introduction}
Reliable transfer of confidential information is becoming more and more 
important. 
The only means, mathematically proven to be secure, is the One
Time Pad \cite{onetimepad}, in which the sender, Alice, and the receiver,
Bob, need to 
share a common secret key. This 
key is used to en- and decode the secret message. 
Quantum key distribution (QKD) is 
known to complement the One Time Pad to a secure system \cite{physworld}. 
Based on the  
non-classical features of quantum mechanics, it provides the distribution
of the key in 
a way, that guarantees the detection of any eavesdropping. Roughly
speaking,
since it 
is not possible to measure an unknown quantum system without modifying it,
an eavesdropper manifests itself by introducing errors 
in the transmission data. After the final
transmission, a suitable subset of data is 
used to estimate the error rate. If the error rate is found
to be below a certain threshold, Alice and Bob can proceed using the
remaining data to establish a secret key by means of error correction and
privacy amplification \cite{errorcorrection}. Otherwise they will decide
that the data are not
secure and start a new data transmission.

During the last years several quantum key distribution  protocols based
on two-level systems
(qubits) have 
been published ~\cite{theoQKD,both}. The connection between the quantum
bit error
rate QBER 
and the maximum amount of 
information the eavesdropper might have attained has been investigated
both in the case of incoherent and coherent (joint measurement of
more qubits) eavesdropping attacks ~\cite{both,eaves}, and the shortening
of the
raw key due
to
error correction and privacy amplification has been calculated \cite{priv}.
There is one interesting thing to note: all work, not only in the case of
quantum key distribution but in 
quantum information theory in generally, has this far mainly been  
based on qubits (for exceptions, see e.g. \cite{mainlybits}). 
The reason is probably easy but not very
scientific:
classical information theory is 
based on bits. Indeed, the enlargement to higher alphabets is of no 
interest in the classical case
since it does not hold any advantage.

From the experimental side,
a number of prototypes, based on qubits, have been
developed, demonstrating 
that QKD not only works 
inside the laboratory, but outside - under real conditions - as well
\cite{expQKD}. 
Thus quantum key distribution together with the One Time Pad, could
already today provide an 
alternative to the traditional public key systems where security is  based 
on computational complexity, which may turn out to be insecure. 
However, using present technology, the prototypes, especially those,
adapted to
"long" transmission distances of tens of kilometers, still 
suffer from low key creation rates of some hundred Hz.
This could be improved by a factor of ten by sending single photon Fock
states instead 
of faint laser pulses (containing only
0.1 photon per pulse), and by increasing the initial pulse rate, however,
the drawback being
more and more severe technical demands.

In this article, we propose a new QKD protocol 
using a larger alphabet and analyze it in terms of key creation rate and 
security against eavesdropping. Surprisingly and in opposition to the
classical treatment of information,
we find some important differences when passing from the two- to the higher
dimensional
case. Since each photon carries more information, 
we find an increase of the flux of information between Alice and Bob  
while keeping the initial pulse rate constant. Beyond and even more
important, a 
first eavesdropping analysis considering the strategy of intercept/resend 
shows advantages of this new protocol with respect to protocols based on
qubits. 

The paper is arranged in the following way: In section 2, we 
introduce our new proposal and investigate 
it in terms of security against eavesdropping. After this theoretical part,
we present an experimental realization (section 3) and 
discuss some further extensions of our idea (section 4). Finally, a 
short conclusion is given in section 5.

\section {Quantum key distribution using larger alphabets}
Higher-dimensional quantum systems have already been 
investigated in order to generalize tests of local realism 
or in the context of the Kochen Specker theorem 
(\cite{Zukowski97,Zukowski99} and references therein).
However, up to now, like its classical counterpart,
quantum information theory has essentially been based on bits (qubits).

Here we investigate the use of higher spins for QKD.
One can imagine a whole variety of new protocols. For example using $n$
non-orthogonal states, where $n$ is the dimension of the space. However
the state identification becomes more and more difficult since it has to
be done with POVM-measurements \cite{povm}. Another possibility is using 
$m$ different 
bases, each with $n$ orthogonal states, where $n$ is the dimension of the
space. In
this
first approach we will limit ourselves to a protocol, using two bases and
four
orthogonal
states per basis.

\subsection{QKD : BB84 in four-dimensions}
The protocol known as the BB84, is
originally intended for two bases, each one with two orthogonal states
(qubits), 
but it may
without difficulty be extended to a four-level system or the so called
quantum quarts (qu-quarts). As in the qubit
case, Alice first chooses in which of two bases
she wants to prepare her state, and second she has to
decide which state to send. In the qu-quart case Alice has to chose
between four different states, whereas in the qubit case she has to
chose between two different states. Each of the two bases are chosen
with equal probability, and each state is again chosen with equal
probability. In other words, in the four-dimensional case, each
of the possible eight states appear with probability $1/8$,   
whereas, in the traditional two-dimensional case, the probability is
$1/4$. 

The first basis can always be chosen arbitrarily as
\begin{eqnarray}
\ket{{\psi}_{\alpha}}, \ket{{\psi}_{\beta}}, \ket{{\psi}_{\gamma}},
\ket{{\psi}_{\delta}},
\label{eq:basis1}
\end{eqnarray}
where the states satisfy $|\braket{{\psi}_{i}}{{\psi}_{j}}|={\delta}_{ij}$
The second basis has to fulfill a certain requirement with respect to the
first basis, namely that $|\braket{{\psi}_{i}}{{\phi}_{j}}|=1/2$. This
requirement makes the protocol symmetric which insures that the
eavesdropper is  not given any advantage. There
are several choice of bases which may fulfill this requirement, but in
the following the basis is assumed to be 
\begin{eqnarray}
\begin{array}{l}
\ket{{\phi}_{\alpha}}=\frac{1}{2}\left( \right.
\ket{{\psi}_{\alpha}}+\ket{{\psi}_{\beta}}+\ket{{\psi}_{\gamma}}+
\ket{{\psi}_{\delta}}
\left. \right)
\\
\ket{{\phi}_{\beta}}=\frac{1}{2}\left( \right.
\ket{{\psi}_{\alpha}}-\ket{{\psi}_{\beta}}+\ket{{\psi}_{\gamma}}- 
\ket{{\psi}_{\delta}}
\left.\right)
\\
\ket{{\phi}_{\gamma}}=\frac{1}{2}\left(\right.
\ket{{\psi}_{\alpha}}-\ket{{\psi}_{\beta}}-\ket{{\psi}_{\gamma}}+
\ket{{\psi}_{\delta}}
\left.\right)
\\
\ket{{\phi}_{\delta}}=\frac{1}{2}\left( \right.
\ket{{\psi}_{\alpha}}+\ket{{\psi}_{\beta}}-\ket{{\psi}_{\gamma}}-
\ket{{\psi}_{\delta}}
\left.\right).
\\
\end{array}
\label{eq:basis2}
\end{eqnarray}
These states  satisfy
$|\braket{{\phi}_{i}}{{\phi}_{j}}|={\delta}_{ij}$. Furthermore the
overlap between
any state from the first ($\psi$-) basis with any state from the second 
($\phi$-) basis is
seen to be $\frac{1}{2}$ as required.

Bob will every time he receives a state chose to measure either in the
$\psi$- or the $\phi$- basis. 
At the end of all the transmissions Alice and
Bob will --- as in the qubit case --- have a public discussion where
they single out the transmission where they have used the same
basis. Since they both make random choices on average $1/2$ of the
transmission have to be discarded. 
If no  eavesdropper is present Alice and Bob will now share
a random string of $\alpha$, $\beta$, $\gamma$ and $\delta$'s, where the
various letters are
taken to be the subscript of the states, i.e. $\ket{{\psi}_{\alpha}}$
and $\ket{{\phi}_{\alpha}}$ are identified as the letter '$\alpha$', etc. 
Thus, if Bob initially made $n$ detections, he ends up with $n/2$ quarts,
which is,
in terms of information contents, equivalent to n bits.
This key can now directly be used to encode a secret
message using the One Time Pad. Note  that the One Time Pad is
not restricted
to bits, but that any alphabet can be used, see e.g.\cite{onetimepad}.

\subsection{Eavesdropping --- Intercept/resend}
During the public discussion Alice and Bob extract a subset of data which
is compared in public. This discussion leads to an estimate of the
error rate induced by the presence of an eavesdropper. The data revealed
during the discussion is afterwards discarded.

The simplest possible eavesdropping strategy is the intercept/resend
strategy, in which  Eve (the eavesdropper) intercepts the transmissions
from Alice
to Bob, performs a measurement and, according to the outcome of her
measurement, she prepares a new state and sends it on to Bob. In the
following only the cases where Alice and Bob use the same basis are
considered, since the ones where they use different bases are discarded
during the public discussion.

Suppose Alice sends the state $\ket{{\psi}_{\alpha}}$. If Eve performs 
her measurement in the $\psi -$basis she will find the state  
$\ket{{\psi}_{\alpha}}$ and she will prepare a new $\ket{{\psi}_{\alpha}}$
state and send it to Bob. Hence Eve introduces no errors and Bob finds
the correct state. If instead Eve measures in  the  $\phi -$basis, she
will with equal probability, $1/4$, find one of the four different 
$\phi -$states and pass it on to Bob. For any of the $\phi-$states Bob
will only find the correct state, $\ket{{\psi}_{\alpha}}$, with
probability $1/4$ ---  which means that with probability $3/4$ he will
get the wrong state, hence an error.

For the following discussions it is convenient to introduce a more
formal measure of information. The relevant information measure here is
Shannon Information \cite{info} which by tradition is measured in terms of
bits. The
Shannon information is for qubits bounded between 0 and 1 bit, since
each qubit can carry one bit of information. Whereas for the qu-quarts
the Shannon information is bounded between 0 and 2 bits since each
qu-quart can carry two bits of information. To obtain 0 bits of course
means obtaining no information, and  1 bit or 2 bits,
respectively, means having full information.  The general form of the
Shannon information is 
\begin{eqnarray}
I_S=\left\{  \begin{array}{l}
1+H(p_1,..,p_n)~~~{\rm for ~qubits}\\
2+H(p_1,..,p_n)~~~{\rm for ~qu-quarts} 
\end{array}\right.
\end{eqnarray}
where $H(p_1,..,p_n)$ is the entropy function  defined as
$H(p_1,..,p_n)=p_1\log_2 p_1 + \cdots +p_n\log_2 p_n$, and $p_1,..p_n$ is
the probability distribution of the possible outcomes $1...n$.

Above it was argued  that the eavesdropper learns correctly half of
the transmissions when using the intercept/resend strategy. The
formal definition of Shannon information leads for the qu-quart case to
\begin{eqnarray}
\begin{array}{lll}
I_S^4& =& \frac{1}{2}
\left[ \right.2+1\log_2 1 +0\log_2 0 +0\log_2 0 +0\log_2 0\left.\right]
+\\
& &\frac{1}{2}
\left[\right. 2+\frac{1}{4}\log_2 \frac{1}{4} +\frac{1}{4}\log_2 \frac{1}{4} 
+\frac{1}{4}\log_2 \frac{1}{4} +\frac{1}{4}\log_2 \frac{1}{4}\left.\right] \\
&=&1
\end{array}
\end{eqnarray}
as expected, since half of the times Eve learns 2 bits and half of the
times she learns 0 bits, leading to an average of 1 bit --- half of 
the transferred information.

Calculating Eve's amount of Shannon information 
for the qubit case leads to
\begin{eqnarray}
\begin{array}{lll}
I_S^2& =& \frac{1}{2}
\left[ \right.1+1\log_2 1 +0\log_2 0 \left.\right] +\frac{1}{2}
\left[\right. 1+\frac{1}{2}\log_2 \frac{1}{2} +\frac{1}{2}\log_2
\frac{1}{2}
\left.\right] \\
&=&\frac{1}{2}
\end{array}
\end{eqnarray}
This means that also in the qubit case the eavesdropper learns on
average only half of the transfered information, however, 
she will introduce a smaller error rate of only one out of four 
transmissions \cite{errorcorrection}. 
To conclude,
the eavesdropper gains the same fraction of information
whether using  qu-quarts or qubits,
however the qu-quarts do hold an advantage for Alice and Bob, since the
eavesdropper introduces a higher error rate --- $3/8$ compared to $1/4$
--- in order to obtain the same
amount of information \cite{infinity}.
From now on, we will
refer to the quantum error rate in the general case as quantum 
transmission error rate QTER.

\subsubsection{The intermediate basis}      
As in the qubit case Eve may also chose to perform her measurement in
what is known as the intermediate basis, instead of using the same
bases as Alice and Bob \cite{errorcorrection}. 
Eavesdropping in the intermediate basis is the simplest example of an
eavesdropping strategy which gives the eavesdropper probabilistic
information.

In the extended BB84 protocol the intermediate
$\theta -$ basis is defined with the following requirements
\begin{eqnarray}
\begin{array}{l}
|\braket{{\theta}_{i}}{{\psi}_{i}}|=
|\braket{{\theta}_{i}}{{\phi}_{i}}|=~~~{\rm max~value}\\
|\braket{{\theta}_{i}}{{\psi}_{j}}|=
|\braket{{\theta}_{i}}{{\phi}_{j}}|=~~~{\rm min~value,~for~all~}i\neq j\\
\end{array}
\end{eqnarray}
The two vectors $\ket{{\psi}_{i}}$ and $\ket{{\phi}_{i}}$ define a
plane. The vector which gives the same maximum overlap, is the one with
the same minimum distance to both of them, which
is 
\begin{eqnarray}
\ket{{\theta}_{i}}= N \left(\ket{{\psi}_{i}} + \ket{{\phi}_{i}}\right)
\end{eqnarray}
where $N$ is the normalization constant. This argument leads to the
following intermediate basis
\begin{eqnarray}
\begin{array}{c}
\ket{{\theta}_{\alpha}}=\frac{1}{2\sqrt{3}}\left( \right.
3\ket{{\psi}_{\alpha}}+\ket{{\psi}_{\beta}}+\ket{{\psi}_{\gamma}}+
\ket{{\psi}_{\delta}}
\left. \right)
\\
\ket{{\theta}_{\beta}}=\frac{1}{2\sqrt{3}}\left( \right.
\ket{{\psi}_{\alpha}}-3\ket{{\psi}_{\beta}}+\ket{{\psi}_{\gamma}}- 
\ket{{\psi}_{\delta}}
\left.\right)
\\
\ket{{\theta}_{\gamma}}=\frac{1}{2\sqrt{3}}\left(\right.
\ket{{\psi}_{\alpha}}-\ket{{\psi}_{\beta}}-3\ket{{\psi}_{\gamma}}+
\ket{{\psi}_{\delta}}
\left.\right)
\\
\ket{{\theta}_{\delta}}=\frac{1}{2\sqrt{3}}\left( \right.
\ket{{\psi}_{\alpha}}+\ket{{\psi}_{\beta}}-\ket{{\psi}_{\gamma}}-
3\ket{{\psi}_{\delta}}
\left.\right)
\\
\end{array}
\label{eq:basis3}
\end{eqnarray}
which satisfy $|\braket{{\theta}_{i}}{{\theta}_{j}}|={\delta}_{ij}$ and
which has the following overlaps 
\begin{eqnarray}
\begin{array}{l}
|\braket{{\theta}_{i}}{{\psi}_{i}}|=
|\braket{{\theta}_{i}}{{\phi}_{i}}|=\frac{3}{2\sqrt{3}}\\
|\braket{{\theta}_{i}}{{\psi}_{j}}|=
|\braket{{\theta}_{i}}{{\phi}_{j}}|=\frac{1}{2\sqrt{3}}\\
\end{array}
\end{eqnarray}

Assume that Alice sends the state $\ket{{\psi}_{\alpha}}$ and Eve
measures in the intermediate basis, i.e. the $\theta -$basis, then she
will find the following outcomes with the corresponding probabilities:
$P({\theta}_{\alpha})=3/4$ and  $P({\theta}_{\beta})= 
P({\theta}_{\gamma})=P({\theta}_{\delta})=1/12$. These probabilities
give Eve the following Shannon information
\begin{eqnarray}
I_S^4=\left( \right.2+{\frac{3}{4}}\log_2 {\frac{3}{4}}+3{\frac{1}{12}}
\log_2{\frac{1}{12}}\left.\right) \approx 0.792
\end{eqnarray}
and she will give rise to the following error rate:
with probability $3/4$ Eve will send to Bob the state 
$\ket{{\theta}_{\alpha}}$ and with probability $1/12$ she will send him 
$\ket{{\theta}_{\beta}}$, $\ket{{\theta}_{\gamma}}$ or 
$\ket{{\theta}_{\delta}}$. Which gives Bob the following probability of
finding the correct state (remember that Alice sent
$\ket{{\psi}_{\alpha}}$, and it is assumed that Bob measures in the
$\psi -$basis) $\frac{3}{4}\frac{3}{4}+ 3(\frac{1}{12}\frac{1}{12})=
\frac{5}{8}$. Which means that even if Eve measures in the intermediate
basis she will introduce the same error rate, namely $3/8$.

Comparing this strategy on the qu-quarts with the equivalent strategy on
qubits; In the qubit case Eve has probability $P(s)=\frac{2-\sqrt{2}}{4} 
\approx 0.146$ for
successfully identifying the state, leading to an amount of
Shannon information of 
\begin{eqnarray}
I_S^2=\left( \right.1+({\scriptstyle \frac{2-\sqrt{2}}{4}})\log_2
({\scriptstyle 
\frac{2-\sqrt{2}}{4}})+(1-{\scriptstyle 
\frac{2-\sqrt{2}}{4}})\log_2(1-{\scriptstyle \frac{2-\sqrt{2}}{4}})
\left.\right)\approx 0.399
\end{eqnarray} 
and an QBER of $1/4$.
To compare the information gained by the eavesdropper in the two
cases, we have to consider how much information she has on the whole
string. Suppose that Alice has sent $n$ qu-quarts, then Eve has $0.792n$ 
bits of information on the whole string. In order to transmit the same
amount of information to Bob using qubits, Alice has to send $2n$ qubits 
since each qubit carries half the amount of information of a qu-quart.
This means that in the qubit case Eve would obtain
$2n\times0.399=0.798n$
bits of information.  
Again this shows that in the case of a larger alphabet the eavesdropper
will introduce a higher error rate --- the same as in the case treated
before (2.2)----
in order to get a comparable amount of
information. The issue of optimal eavesdropping on the higher alphabet
will be discussed in a forthcoming paper \cite{neweve}.

\subsection{Mapping onto a two-dimensional key}
Classically a larger alphabet like the one used here, may simply be viewed
as an encoding of bits, for example $\alpha=00$, $\beta=01$, $\gamma=10$
and $\delta=11$. Alice and Bob can also in this case chose to
view the higher alphabet as a simple encoding of bits. However,  the
following example shows that they have to be careful about when they
perform the translation. 
Suppose that the eavesdropper has used the intermediate
basis\footnote{The 
same can not be illustrated considering the
eavesdropping strategy where Eve use the same basis as Alice and Bob,
since in that case she has either full information about the quart 
sent by Alice or no information at all.}, then she
will have obtained each quart correctly with probability $3/4$.
This means
that on average Eve will have $3$ out of $4$ quarts correctly ---
however she does not know which ones she has correctly and which ones are
wrong. Now suppose that Alice has sent the following string of
$\alpha$, $\beta$, $\gamma$ and $\delta$:
\begin{eqnarray}
\begin{array}{rcl}
{\rm Alice}&:&~\alpha~\delta~\beta~\alpha~\gamma~\delta~
\delta~\beta~\gamma~\alpha~
\beta~\delta...\nonumber
\end{array}
\end{eqnarray}
but that Eve has the string 
\begin{eqnarray}
\begin{array}{rcl}
{\rm Eve}&:&~\alpha~\delta~\mbox{\boldmath$\gamma$}~\alpha~\gamma~
\mbox{\boldmath$\beta$}~
\delta~\beta~\gamma~\alpha~
\mbox{\boldmath$\alpha$}~\delta...\nonumber
\end{array}
\end{eqnarray}
It is easily seen that $3$ out of $12$ are wrong in Eve's string  or
that she has $9/12=3/4$ correct.

Assume now that Alice and Bob want to map the 4-dimensional key onto a
binary one. Using the above given example, Alice's  new string reads
\begin{eqnarray}
\begin{array}{rcl}
{\rm Alice}&:&~00~11~01~00~10~11~11~01~10~00~01~11...\nonumber
\end{array}
\end{eqnarray}
Doing the same with her sequence of quarts, Eve ends up with
\begin{eqnarray}
\begin{array}{rcl}
{\rm Eve}&:&~00~11~{\bf 10}~00~10~{\bf 0}1~11~01~10~00~0{\bf
0}~11...\nonumber
\end{array}
\end{eqnarray}
Notice that here Eve has $4/24=1/6$ bits wrong or $5$ out of $6$ bits
correct. This is due to the fact that the errors occurring in Eve's
string are no longer
independent, but depend on each other in the respective blocks.
Beyond, the
processes of error correction and privacy amplification do not apply to
this case. This means
that Alice and Bob have to perform this process in the higher alphabet,
and  
only perform the translation to bits at the very end when the eavesdropper
has no information on the string shared between them.

It is important to realize that Alice and Bob should not even discuss how
the translation should be done until after error correction and privacy
amplification, since this information may give an advantage
to
the eavesdropper. Assume that Alice and Bob before starting the
transmissions of qu-quarts have decided for the bit-encoding which is
given
above. In designing the optimal eavesdropping strategy, Eve may use this
knowledge to give different weight to the various states. Suppose, for
example, Alice
sends an $\alpha$. Eve will with the highest possible probability try to
identify that Alice sent an $\alpha$, since in that case she has learned
both bits correctly. Failing to make the correct identification, and
instead obtaining $\beta$ or $\gamma$ will however still give her one bit
correct, whereas finding $\delta$ gives her only errors. As a consequence
Eve will give more weight to $\beta$ and $\gamma$, than to $\delta$.
However, this
is very different from eavesdropping on the higher alphabet, where
obtaining
$\alpha$ is correct, but any other letter $\beta$, $\gamma$ or
$\delta$ is equally wrong.

\section{Experimental realization}
In the following, the given states are given physical meaning and 
a possible experimental realization for a four-letter
alphabet is presented. 
It is important to notice that
generalization to arbitrarily large alphabets 
is in principle possible.

Alice is in possession of an apparatus (see fig.1) that allows her to route
an incoming photon 
(or faint laser pulse) emitted at time $t_0$ to one of four different delay
lines. 
This task can be accomplished by using an optical switch. Using another
switch, 
the light traveling via the chosen delay line is then injected 
(at times $t_{\alpha}$,..,$t_{\delta}$) into the output port of the device. 
This apparatus thus allows Alice to create single photons 
in four different time slots which we identify
with the states $\ket{\psi_{\alpha}},\cdots,\ket{\psi_{\delta}}$.  
To distinguish the four states of the $\psi$- or time-basis, 
it suffices to measure the arrival time of the photons 
with respect to $t_0$. Bobs analyzer thus simply consists of a 
photon detector and a fast clock.

In order to create one of the four states belonging to the $\phi$- or
energy-basis,
Alice has to
prepare a coherent superposition of the four emission times with
appropriate phase 
differences. Hence, the first switch in the preparation device has to be
replaced by a 
symmetric 1 x 4 optical coupler. Using the coupler depicted in Figure 2
\cite{Zukowski97}
and phases $\alpha_A,\cdots,\delta_A$, it is not difficult to show that 
it is possible to create the desired states $\ket{\phi_{\alpha}},\cdots,
\ket{\phi_{\delta}}$. 
For instance, 
choosing $\alpha_A$= 0, $\beta_A$=$\pi$/2, $\gamma_A$= 0 and $\delta_A$=
$-\pi$/2
leads (neglecting an unimportant overall phase)
to creation of $\ket{\phi_{\alpha}}$ \cite{differentinputports}. 
In order 
to distinguish the four states of the energy basis, Bob has to have a device 
that coherently recombines the pulses arriving at times
$t_{\alpha},\cdots,t_{\delta}$ 
and then makes them interfere in such a way, that each state leads 
to completely 
constructive interference in specific output port, hence to a 
detection by a different detector.
The device thus consists of an optical switch that routes the pulse
arriving first to a 
long delay line, the pulse arriving second to a shorter one etc. 
in a way, that the delay difference 
introduced by Alice will be exactly compensated. Using the already
mentioned 4 x 4 coupler and 
phase settings $\alpha_B =0,\beta_B=-\pi/2, \gamma_B=0$ and
$\delta_B=\pi/2$, 
it is straightforward to show that each state indeed leads to detection in a 
different detector 
(i.e., $\ket{\phi_{\alpha}}$ leads to detection in $D_{\alpha}$ etc).

\section{Discussion and experimental extensions}

As stated in the introduction, 
one of the motivations for considering higher dimensional systems for QKD
is the increase of information per photon. 
This leads after error correction and privacy amplification of 
the high-dimensional (more-then-two-dimensional) key
to a larger binary key.
Beyond, the important point of our proposal is that the mentioned 
speedup goes along with an increasing quantum transmission error rate 
introduced by 
eavesdropping. Unfortunately, this advantage of easier 
detection of an unlegitimized third person is 
somewhat hidden by a higher QTER introduced by Alice and Bob themselves. 
That is, the larger number of 
simultaneously active (and noisy) detectors engenders a higher probability
that one detector 
sees a dark count while a photon is expected to arrive. 
However, we believe that
as long as those errors are small, QKD using higher 
alphabets could still be advantageous compared to two-level systems. 

Attention should be drawn to several interesting extensions of
this proposal.
First of all, similar to QKD schemes based on two-level systems, it is
possible to find a 
"plug$\&$play" system \cite{plug&play} using time multiplexed
interferometry to realize
systems of higher 
dimensions as well, the advantage being that there is no need to equalize 
the path differences of different interferometers. 
Second, our proposal can easily be extended to photon correlation
experiments as well (for proposals in the domain of fundamental physics,
see i.e.
\cite{Zukowski97,Zukowski99}).
To give an example, 
higher dimensional secret sharing \cite{secretsharing}
could be realized 
in the following way.
One could pump a nonlinear crystal with a laser pulse, having
traveled via one 
out of n delay lines, or a coherent superposition of n delay lines,
respectively. Similar to the here treated case, the 
created photon pair would then be described either in a time, or an energy
basis.
After having separated the two photons, each one is then analyzed  
in one of the two bases (in this context, see
\cite{newsource, cryptonewsource}).

\section{Conclusion}

We proposed to enlarge the dimensions of quantum systems for quantum key
distribution and analyzed a protocol based on four orthogonal states in
two 
different bases in terms of information transfer and 
introduced errors by an intercept/resend eavesdropping attack. 
Since every particle now carries more information, we find an increased
flux of information
which can be turned into an increased binary key creation rate. Beyond, the
quantum transmission error rate introduced by 
an eavesdropper for a given amount of acquired information is much higher
then in 
the qubit case. Furthermore we proposed an experimental realization using
an interferometric setup.
Even if
the quantum transmission error rate introduced 
by the noise of the detectors is higher than in the qubit case, the new
protocol will still
be advantageous compared to two-level systems as long as these
errors are small.

Besides these more practical considerations,
the most important point of our proposal is that, in opposition to its 
classical counterpart, 
quantum information theory, at least quantum key distribution, 
changes when passing from two dimensional to higher dimensional systems. 
We thus believe that it might be interesting to consider other 
applications from this the point of view 
as well.

\section*{Acknowledgement}
We would like to thank Nicolas Gisin for many stimulating discussions, and
Nobert Lutkenhaus for helpful comments. 
H.B.-P. is supported by the Danish National Science
Research Council (grant no. 9601645). This work was also supported by the
European IST project EQUIP.


\begin{figure}
\infig{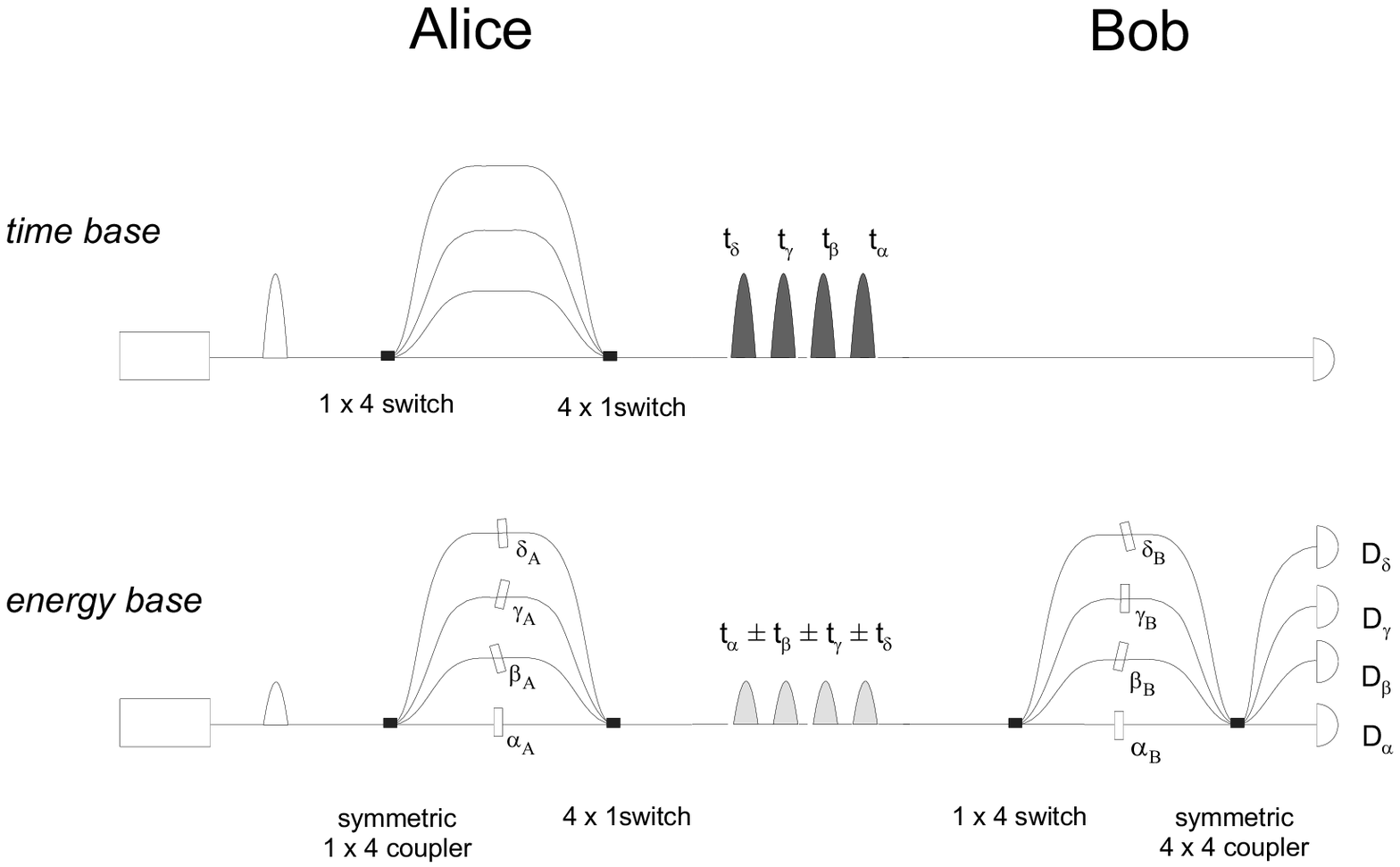}{0.85\columnwidth}
\caption{Schematical setup for four-letter quantum key distribution. 
See text for detailed description.}
\label{fig1} 

\vspace{2cm}

\infig{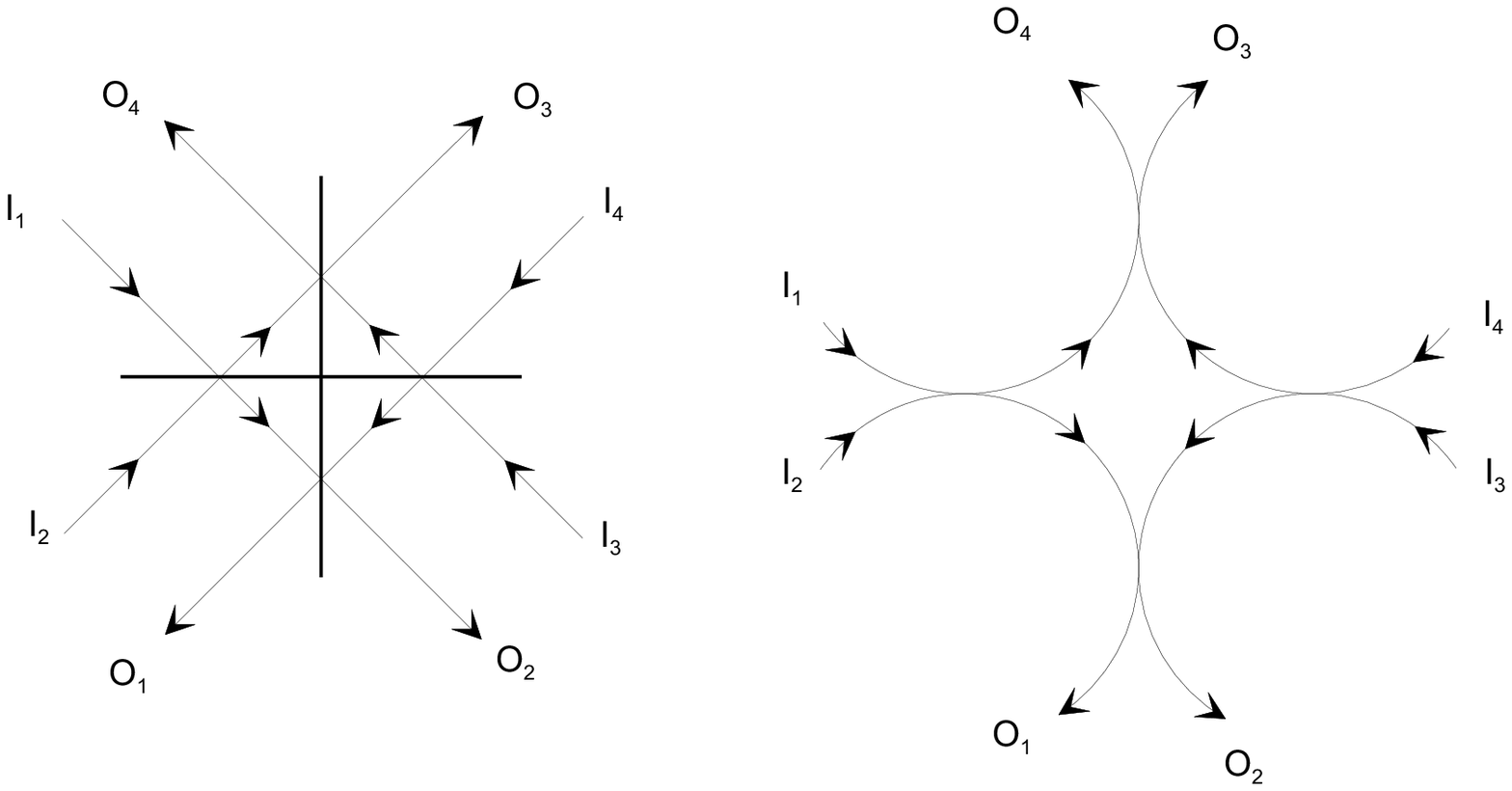}{0.85\columnwidth}
\caption{Bulk- and fiber optical realization of a symmetric 4x4 coupler.}
\label{fig2}

\end{figure}


\begin{thebibliography}{99}

\bibitem{onetimepad}See e.g. D. Welsh, Codes and Cryptography, Oxford
Science Publication, Clarendon Press, Oxford, 1988

\bibitem{physworld}Physics World, March 1998, special issue in quantum
communication, 
including an article by W. Tittel, G.Ribordy, and N.Gisin on quantum
cryptography.

\bibitem{errorcorrection}C. Bennett, F. Bessette, G. Brassard,
L. Salvail, J. Smolin, J. Cryptology (1992)5: 3-28

\bibitem{theoQKD}C.H. Bennett and G. Brassard, in Proceedings of IEEE
International 
Conference on Computers, Systems and Signal Processing, Bangalore, India 
(IEEE, New York, 1984), p.175;
A.K. Ekert, Phys. Rev. Lett. {\bf67}, 661 (1991);
C.H. Bennett, Phys. Rev. Lett. {\bf68}, 3121 (1992);
B. Huttner, N. Imoto, N. Gisin, and T. Mor, Phys. Rev. A {\bf51}, 1863
(1995);
 

\bibitem{both}D. Bruss, Phys. Rev. Lett {\bf 81}, 3018 (1998); 
H. Bechmann-Pasquinucci and N. Gisin, Phys. Rev. A {\bf59}, 4238 (1999).



\bibitem{eaves}C. Fuchs, N. Gisin, R. B. Griffiths, C. S. Niu and A. Peres
Phys. Rev. A {\bf 56}, 1163 (19997); I. Cirac and N. Gisin, Phys. Lett. A
{\bf
229}, 1 (1997); 

\bibitem{priv}C. Bennett, G. Brassard, C. Crepeau and  U. Maurer, IEEE
Trans. Inf. Theory {\bf 41}, 1915 (1995); N. Lutkenhaus, Phys. Rev. A {\bf
54}, 97 (1996)

\bibitem{mainlybits}E. Knill, quant-ph/9608048, quant-ph/9608049;
D. Gottesman, quant-ph/9802007

\bibitem{expQKD}H. Zbinden, H. Bechmann-Pasquinucci, N. Gisin, and G. Ribordy.
Appl. Phys. B {\bf67}, 743 (1998). 


\bibitem{Zukowski97}M. Zukowski, A. Zeilinger, and M. A. Horne, Phys.Rev.A
{\bf55}, 2564 (1997).

\bibitem{Zukowski99}M. Zukowski and D. Kaszlikowski, Phys. Rev. A {\bf59},
3200 (1999).

\bibitem{povm}A. Peres, Quantum Theory: Concepts and Methods, Kluwer
Academic Publishers, Dordrecht, 1995

\bibitem{info}T. Cover and J. Thomas, Elements of Information Theory,
Wiley Series in Telecommunications, John Wiley \& Sons, Inc. New York 1991.

\bibitem{infinity}To generalize to infinitly high dimensions : 
Eve will correctly learn half of the transmissions, but will introduce 
an error rate of $1/2$.

\bibitem{neweve}H. Bechmann-Pasquinucci et al. in preparation

\bibitem{differentinputports} Another experimental realization might be to 
keep the phases in the interferometer 
stable and to trigger different input ports in order to generate the
different states.

\bibitem{plug&play}A. Muller, T. Herzog, B. Huttner, W. Tittel, H. 
Zbinden, and
N. Gisin,
Appl. Phys. Lett {\bf70}, 793 (1997). 
G. Ribordy, J.-D. Gautier, N. Gisin, O. Guinnard, and H. Zbinden, 
Electr.Lett. {\bf34} (22), 2116 (1998).

\bibitem{secretsharing}A. Karlsson, M. Koashi, and N. Imoto, Phys.Rev
A{\bf59}, 162 (1999);
 M. Hillery, V. Buzek, and A. Berthiaume, quant-ph/9806063.

\bibitem{newsource}J. Brendel, N. Gisin, W. Tittel, and H. Zbinden,
Phys.Rev.Lett. {\bf82} (1999).

\bibitem{cryptonewsource} W. Tittel et al. In preparation.

\end{thebibliography}
\end{document}